# Power spectrum with auxiliary fields in de Sitter space

M. Mohsenzadeh,[1,][*] M.R. Tanhayi,[2,][†] and E. Yusofi[3,][‡]

[1]*Department of Physics, Qom Branch, Islamic Azad University, Qom, Iran*
[2]*Department of Physics, Islamic Azad University, Central Tehran Branch, Tehran, Iran*
[3]*Department of Physics, Science and Research Ayatollah Amoli Branch,
Islamic Azad University, Amol, Mazandaran, Iran*
(Dated: March 28, 2014)

We use the auxiliary fields and (excited-) de Sitter solutions to study the standard power spectrum of primordial fluctuations of a scalar field in the early universe. The auxiliary fields are the negative norm solutions of the field equation and as it is shown, with a fixed boundary condition, utilizing these states results in a finite power spectrum without any infinity. The power spectrum is determined by the de Sitter solutions up to some corrections and the space-time symmetry is not broken in this point of view. The modulation to the power spectrum is of order $(\frac{H}{\Lambda})^2$, where $H$ is the Hubble parameter and $\Lambda$ is the energy scale, e.g., Planck scale.

## I. INTRODUCTION AND MOTIVATION

The cosmic microwave background (CMB) which is the thermal radiation decoupled from the cosmic gas after the Big Bang, provides a snapshot of the early universe that has been studied in many papers ([1] and references therein). There is an anisotropy in the temperature map of CMB that could come from the primordial perturbations of quantum fields at the very early universe which indicates the quantum origin of the universe. It may also be related to physics beyond the Planck scale as long as the fluctuations start out with a linear size much smaller than the Planck scale [2]. Such an anisotropy affects the spectrum of the perturbations and in some models it is argued that the order of such modifications on the size of spectrum cannot be larger than $\mathcal{O}(\frac{H}{\Lambda})^2$ (for a good review see [3]). Analysing the data coming from Planck satellite can help one to provide significant constraints on the theoretical aspect of inflation and also pose an important challenge to competing scenarios for the origin of the initial perturbations.

In this paper we use the auxiliary modes [such modes have negative norm according to the proposed inner product] to study such a correction to the spectrum and the key point in our approach is that the infinity in calculating the correlation functions does not appear. It is well known that the inflation can be described in approximate de Sitter (dS) space-time [4], thus, the study of dS and quantum theory of fields in this background is well motivated. For example, exploring the possible relations between the dS symmetries and the bispectrum of the fluctuations are well studied in [5] in order to set constraints on the initial fluctuations due to the dS symmetry, especially, how scale transformations and special conformal symmetries constrain the correlation functions. The scalar field in dS background is important because most of inflationary models are theorized using the scalar field [6]. It is proved that due to the famous zero-mode problem, quantization of the massless fields in dS can not be done covariantly [7, 8]. In other words a proper dS invariant vacuum state can not be constructed with only positive norms and one needs auxiliary fields and a Gupta-Bleuler type construction based on indefinite metric field quantization method (Krein space instead of Hilbert space) is actually needed [9]. This means that for a covariant quantization one should use the negative norm solutions or auxiliary states. Consideration of the negative norm states was first studied by Dirac in 1942 [10]; Gupta and Bleuler in 1950 used such states to remove infrared divergence in QED [11]. Carrying out the quantization in Krein space has followed in some papers: the vacuum energy vanishes and infinite term does not appear in the calculation of the expectation value of $T_{\mu\nu}$ [12], in the interaction QFT this method automatically removes the singular behaviors (ultraviolet) of Green's functions [13, 14]. In the present work, we use this method to study the power spectrum of the scalar field in dS background by considering an approximate dS solution and as it is shown the singularity of the power spectrum is removed automatically. In [15], through the Krein space quantization method, the spectrum of gravitational waves produced during the inflation in slow roll approximation has been studied.

---

[*]Electronic address: mohsenzadeh@qom-iau.ac.ir
[†]Electronic address: m_tanhayi@iauctb.ac.ir
[‡]Electronic address: e.yusofi@iauamol.ac.ir

The layout of paper is as follows: In Section 2, we briefly recall the definition of standard power spectrum. Indefinite metric field quantization method or Krein space method is used to calculate the power spectrum in section 3. A conclusion and an outlook is given in the final section.

## II. POWER SPECTRUM: BASIC SET-UP

The following metric of dS is used to describe the universe during the inflation:

$$ds^2 = dt^2 - a(t)^2 d\mathbf{x}^2 = a(\eta)^2 (d\eta^2 - d\mathbf{x}^2), \tag{II.1}$$

where for the conformal time $\eta$ the scale factor is defined by $a(\eta) = -\frac{1}{H\eta}$. There are some models of inflation but the single field inflation in which a minimally coupled scalar field (inflaton) in dS background, is usually studied in the literature. The action is given by:

$$S = \frac{1}{2} \int d^4x \sqrt{-g} \Big( R - (\nabla \phi)^2 - m^2 \phi^2 \Big),$$

where $M_{Planck}^{-2} = 8\pi G = 1$. The corresponding massless field equation is given by the usual Klein-Gordon equation which takes the following form in conformal coordinate:

$$\phi'' + 2\frac{a'}{a}\phi' - \nabla^2 \phi = 0, \tag{II.2}$$

where the prime is the derivative with respect to conformal time. In Fourier space by substituting $\phi_k(\eta) = \frac{1}{a} u_k(\eta)$, the inflaton field equation turns to

$$u_k'' + (k^2 - \frac{a''}{a}) u_k = 0. \tag{II.3}$$

One can quantize $\phi$ by considering the general solution of (II.3) as:

$$u_k = \frac{\sqrt{\pi \eta}}{2} \Big( D_- H_{3/2}^{(1)}(k\eta) + D_+ H_{3/2}^{(2)}(k\eta) \Big), \tag{II.4}$$

where $H_\nu$ are the Hankel functions [16]. The quantum mode $\hat{u}$, becomes

$$\hat{u}(x,\eta) = \int \frac{d^3 \mathbf{k}}{(2\pi)^3} \Big( \hat{a}_\mathbf{k} u_k(\eta) e^{i\mathbf{k}\cdot\mathbf{x}} + \hat{a}_\mathbf{k}^\dagger u_k^*(\eta) e^{-i\mathbf{k}\cdot\mathbf{x}} \Big), \tag{II.5}$$

where $\hat{a}_\mathbf{k}$ and $\hat{a}_\mathbf{k}^\dagger$ are the annihilation and creation operators, respectively. The degrees of freedom can be indeed fixed by imposing the flat limit condition namely at the early time ($\eta \to -\infty$ or equivalently $|k\eta| \gg 1$ or $k \gg aH$), the corresponding vacuum should match the flat vacuum state [17]. We will get back to this in the next section.

In the linear evolution schema of universe, the density perturbations are encoded by the correlation functions, where for a Gaussian distribution these are completely specified by the spectrum of density fluctuations. In the single field inflation, the scalar field (including both inflaton and associated spatial curvature) and tensor field fluctuations are usually studied [17]. The spectrum of fluctuations is indeed the smoothed modulus-squared of its Fourier coefficient and is defined by [18]:

$$P_\chi \equiv (Lk/2\pi)^3 4\pi \langle |\chi_\mathbf{k}|^2 \rangle, \tag{II.6}$$

$\chi$ stands for any kind of perturbations and $L$ is the size of normalization box for the Fourier expansion, noting that the average is taken over a small region of $k$-space.

Tensor mode fluctuations or the primordial gravitational waves come from the metric fluctuation and one can correspond a free massless scalar field for both independent polarizations of metric fluctuations [19], and approximately one can deduce that their equations of motion are somehow similar to Eq. (II.3). Therefore, it is expected that aside from a series of coefficients, power spectrum for gravitational waves is proportional to its scalar field counterpart [20].

## A. Power spectrum in Hilbert space

In the case of de Sitter metric ($H = const.$ and $\frac{a''}{a} = \frac{2}{\eta^2}$) the exact solution of (II.3) is as follows:

$$u_k = \frac{A_k}{\sqrt{2k}}(1 - \frac{i}{k\eta})e^{-ik\eta} + \frac{B_k}{\sqrt{2k}}(1 + \frac{i}{k\eta})e^{ik\eta}, \tag{II.7}$$

where $A_k$ and $B_k$ are Bogoliubov coefficients. Because there is no time-like Killing vector in dS, thus in general, a set of vacua (labelled by $\alpha$) is used. In other words, the free parameters $A_k$ and $B_k$ characterize the non-uniqueness of the mode functions.[1] However, imposing an initial condition at the very early time together with the normalization condition leads to the Bunch-Davies vacuum in which one sets $A_k = 1$ and $B_k = 0$

$$u_k^{B.D} = \frac{1}{\sqrt{2k}}(1 - \frac{i}{k\eta})e^{-ik\eta}. \tag{II.8}$$

In other words in the limit of $\eta \to -\infty$ for a fixed mode $k$, a scalar field can be well described by quantum field theory in flat space-time, as long as in this limit, the modes can be situated deep in the Hubble horizon and consequently the curvature effect is negligible. This is the initial boundary condition, and the vacuum of the theory coincides with the flat space vacuum and the positive frequency modes are deduced from the Hankel functions (II.8), the resulting vacuum is the thermal or Euclidean vacuum [22].

For a given mode $u_k$, the two-point function in Hilbert space is defined by:

$$\langle \phi^2 \rangle = \frac{1}{(2\pi)^3} \int \frac{|u_k|^2}{a^2} d^3k. \tag{II.9}$$

---

[1] Basically in Minkowski space for a free scalar field, one can find an essentially unique vacuum state which is Poincaré invariant. However, there is not such an unique vacuum in de Sitter space, in fact for a free massive scalar field, a set of vacua pose the invariance under the isometries of de Sitter space [7, 21]. These sates are obtained from (II.7) by considering $A_k, B_k \neq 0$ and the resulting vacua can be parameterized by a single complex number say as $\alpha$ and are usually named $\alpha$-vacua. These vacua have some special features such as a mixture of positive and negative frequency modes at short- distances.

Then from (II.8) and (II.9) one can write:

$$\langle\phi^2\rangle = \frac{1}{(2\pi)^3}\int d^3k\Big[\frac{1}{2ka^2} + \frac{H^2}{2k^3}\Big]. \tag{II.10}$$

The first term is the usual contribution from vacuum fluctuations in Minkowski space-time that can be eliminated after the renormalization [23], then the power spectrum for the scalar field fluctuations is calculated as [18, 19]:

$$P_\phi(k) = \Big(\frac{H}{2\pi}\Big)^2. \tag{II.11}$$

In reference [24], for a different initial condition (namely for $\alpha$-vacua), by considering the trans-Planckian effect that appears as a fixed scale, the fluctuation spectrum has been obtained as follows:

$$P_\phi = \Big(\frac{H}{2\pi}\Big)^2\Big(1 - \frac{H}{\Lambda}\sin(\frac{2\Lambda}{H})\Big). \tag{II.12}$$

Note that for a given $k$ a finite $\eta_0$ is chosen in which the physical momentum corresponding to $k$ is given by some fixed scale $\Lambda$, where $\eta_0 = -\frac{\Lambda}{Hk}$ has a finite value and $\Lambda$ is the energy scale, e.g., Planck scale. This is a scale-dependent power spectrum and corrections are of order $\frac{H}{\Lambda}$. At the next section the power spectrum is considered in Krein space.

### III. POWER SPECTRUM IN KREIN SPACE

As it is known in flat space-time, the vacuum expectation value of the energy-momentum tensor diverges and the divergency is removed by the means of normal ordering procedure. However in curved space-time, the following remedy is usually used (equation (4.5) in [27]),

$$\langle\Omega|:T_{\mu\nu}:|\Omega\rangle = \langle\Omega|T_{\mu\nu}|\Omega\rangle - \langle0|T_{\mu\nu}|0\rangle, \tag{III.13}$$

where $|\Omega\rangle$ is the vacuum of the theory and $|0\rangle$ stands for the flat vacuum state. Note that the mines sign at the above equation might be interpreted as the effect of the background solutions. In this renormalization procedure the vacuum is defined globally while the singularities are removed locally. Indeed the background solutions are no longer solutions of the wave equation in curved space-time, thus it vividly breaks the symmetry. But the symmetry would be preserved if divergences are removed by the quantities which are defined globally. This interpretation of removing infinity resembles Krein space approach, where the renormalization procedure is accomplished by the help of the negative norm solutions of the wave equation and then the minus sign in (III.13) appears because of the auxiliary solutions.

Krein space is the generalization of the Hilbert space in which both negative and positive norm states are present in its construction. Formally, it is defined by $\mathcal{K} = \mathcal{H}_+ \oplus \mathcal{H}_-$, where $\mathcal{H}_-(=\mathcal{H}_+^\star)$ is the 'anti-Hilbert' space. It is proved that after making use of such negative norm states some infinities of the theory are removed. Actually in the Krein methodology, utilizing the negative norm states (with their own ladder operators which are independent of the positive norm states), reduces the singularity of Green-functions [9, 12]. But one might wonder about the instability of the vacuum and also the unitarity of the theory. Introducing an *ansatz*, would address such problems: the elements of the $S$-matrix (probability amplitude) are defined as follows [25, 26]:

$$S_{if} = \frac{\langle in|out\rangle}{\langle 0,in|0,out\rangle}, \tag{III.14}$$

where the states at dominator are the physical states, this guarantees negatively norm states only appear in the internal lines of the Feynman diagrams.

In Krein space, the two-point function is then defined by:

$$\langle \Omega^{Krein}|: \phi^2 :|\Omega^{Krein}\rangle = \langle \phi^2 \rangle_P + \langle \phi^2 \rangle_N \tag{III.15}$$

where the subscript $P$, $(N)$ stands for the positive (negative) norm solutions. In the language of (III.13) this technique means that one removes the effect of the background (flat space in that case) solutions. To illustrate this point let us take the the Bunch-Davies mode (II.8) and calculate the spectrum with the auxiliary modes in flat space as a background (it appears by $u_k^{BG}$ below), then we have:

$$u_k^{dS} = \frac{1}{\sqrt{2k}}(1 - \frac{i}{k\eta})e^{-ik\eta} \quad , \quad u_k^{BG} = \frac{1}{\sqrt{2k}}e^{ik\eta}. \tag{III.16}$$

According to (III.15) and after doing some calculations one finds:

$$\langle \phi^2 \rangle = \frac{1}{(2\pi)^3}\int d^3k \Big(\frac{1}{2ka^2} + \frac{H^2}{2k^3}\Big) - \frac{1}{(2\pi)^3}\int \frac{d^3k}{2ka^2}$$

$$= \frac{1}{2\pi^2}\int \frac{dk}{k}\Big(\frac{H^2}{2}\Big). \tag{III.17}$$

Note that the power spectrum in this case is obtained as $(\frac{H}{2\pi})^2$ which is the same as (II.11). In the case of $\alpha$-vacua, one can write:

$$\langle \phi^2 \rangle = \frac{1}{(2\pi)^3}\int d^3k \Big(\frac{H^2}{2k^3} + \frac{1}{2ka^2} - \frac{H^3}{2\Lambda k^3}\sin(\frac{2\Lambda}{H})\Big) - \frac{1}{(2\pi)^3}\int \frac{d^3k}{2ka^2}$$

$$= \frac{1}{2\pi^2}\int \frac{dk}{k}\Big(\frac{H^2}{2} - \frac{H^3}{2\Lambda}\sin(\frac{2\Lambda}{H})\Big), \tag{III.18}$$

where the power spectrum is also similar to (II.12), noting that the infinity does not appear in these calculations.

Now let us consider an approximate dS solution. Since the inflation takes place in (an approximate) de Sitter space, basically in this high energy area of very early universe with varying $H$, finding a proper mode is difficult. We offer an excited-de Sitter solution as the fundamental mode during the inflation that asymptotically approaches to dS solutions. Such an approximate mode might be obtained by expanding the Hankel function in (II.4) up to its third term [28] and then one can write:

$$u_k \simeq \frac{1}{\sqrt{2k}}\Big(1 - \frac{i}{k\eta} - \frac{1}{2}(\frac{i}{k\eta})^2\Big)e^{-ik\eta}. \tag{III.19}$$

Then the auxiliary or background modes are chosen as:

$$u_k^{BG} = \frac{1}{\sqrt{2k}}(1 + \frac{i}{k\eta})e^{ik\eta}, \tag{III.20}$$

noting that according to the proposed *ansatz*, the negative norm solutions are not affected by the boundary conditions and they only play a renormalizer role in the calculation of power spectrum. After doing some straightforward algebra, one obtains:

$$\langle \phi^2 \rangle = \frac{1}{(2\pi)^3}\int d^3k[\frac{1}{2ka^2} + \frac{H^2}{k^3} + \frac{a^2 H^4}{8k^5}] - \frac{1}{(2\pi)^3}\int d^3k[\frac{1}{2ka^2} + \frac{H^2}{2k^3}]$$

$$= \frac{1}{2\pi^2} \int \frac{dk}{k}\left(\frac{H^2}{2} + \frac{a^2 H^4}{8k^2}\right). \tag{III.21}$$

It is worth noting that if one carries out the quantization at finite wavelength, rather than fully in the ultraviolet (i.e. Bunch Davies) limit, and after substituting $k = ap$ and $p = \Lambda$, the power spectrum becomes

$$P_\phi(k) = \left(\frac{H}{2\pi}\right)^2 \left(1 + \frac{1}{4}\left(\frac{H}{\Lambda}\right)^2\right), \tag{III.22}$$

which is scale-dependent and the correction is of order $\left(\frac{H}{\Lambda}\right)^2$. Note that in [16, 29], similar correction has been obtained.

## IV. CONCLUSIONS

In this paper, we used the auxiliary field to calculate the power spectrum. Quantum fields that contain such states are defined in Krein space. Krein space is built by enlarging the Hilbert space by adding negative norm states. This mathematical approach provides some interesting results which are in agreement with their (Hilbert space) quantum field theory's counterparts e.g., in calculating the vacuum expectation value of the energy-momentum tensor one obtains finite result automatically, or in the Casimir effect of the scalar field the regularized form can be obtained within this formalism [30]. Such auxiliary states are only utilized as a mathematical tool and are not affected by the boundary conditions. It means that in the Feynman diagrams they only appear at the internal legs in the disconnected parts of the diagrams. This also guarantees the unitarity of the theory in studying the $S$-matrix elements [26].

Pursuing this approach, the power spectrum of the inflaton was calculated, it was shown that the results are similar to the previous works, however, in our calculation the infinity does not appear. This inspires some kind of the renormalization, noting that the theory becomes finite itself. In the case of dS background, slightly deviation of the exact solution by expanding the Hankel function for the quantum mode in dS before quantization leads to a correction to the power spectrum which is of order $\left(\frac{H}{\Lambda}\right)^2$. This is similar to many calculations of trans-Planckian modulations to the power spectrum but in our calculations the infinity does not appear. On the other hand, the symmetry of curved space-time has been preserved and the obtained spectrum was scale dependent.

**Acknowlegements**: The authors would like to thank M.V. Takook for useful comments. This work has been supported by the Islamic Azad University-Qom Branch, Qom, Iran.

---

[1] D. N. Spergel et al., Astrophys. J. Suppl. 170: 377, (2007) [astro-ph/0603449].
[2] U. H. Danielsson, JHEP 0207, 040 (2002) [hep-th/0205227].
[3] R. Brandenberger and J. Martin [astro-ph: 1211.6753]
[4] A. D. Linde, Phys. Lett. B 108, 389 (1982); Rept. Prog. Phys. 47, 925 (1984);
    *ibid* Inflationary Cosmology, Lect. Notes Phys. 738 (2008) [arXiv:0705.0164].
[5] J. M. Maldacena and G. L. Pimentel, JHEP 1109: 045, (2011) [arXiv:1104.2846].
[6] C. Wetterich, Nucl. Phys., B 302: 668, (1988).
[7] B. Allen, Phys. Rev. D 32, 3136 (1985).
[8] B. Allen and A. Folacci, Phys. Rev. D 35, 3771 (1987).
[9] J. P. Gazeau, J. Renaud and M. V. Takook, Class. Quant. Grav. 17, 1415 (2000) [gr-qc/9904023].
[10] P. A. M. Dirac, Proc. Roy. Soc. A 180, 1 (1942).

[11] S. N. Gupta, Proc. Phys. Soc. A 63, 681 (1950).
[12] M. V. Takook and S. Rouhani [gr-qc/1208.5562v1].
[13] M. V. Takook, Int. J. Mod. Phys. E 11, 509 (2002) [gr-qc/0006019];
S. Rouhani and M. V. Takook, Euro. Phys. lett. 68, 15 (2004) [gr-qc/0409120].
[14] A. Refaei, M. V. Takook, Phys. Lett. B 704, 326 (2011) [gr-qc/1109.2693].
[15] M. Mohsenzadeh, A. Sojasi and E. Yusofi, Mod. Phys. Lett. A 26, 2697 (2011) [gr-qc/1202.4975].
[16] N. Kaloper, M. Kleban, A. Lawrence and S. Shenker, Phys. Rev. D 66, 123510 (2002) [hep-th/0201158].
[17] D. Baumann, TASI (2009) [hep-th/0907.5424].
[18] A. R. Liddle, and D. H. Lyth, Phys. Rep. 231, 1 (1993).
[19] A. A. Starobinsky, JETP Lett. 30, 683 (1979).
[20] E. W. Kolbe and M. S. Turner, The Early Universe (Addison-Wesley, New York,1990).
[21] E. Mottola, Phys. Rev. D 31, 754 (1985).
[22] T. S. Bunch and P. C. W. Davies, Proc. R. Soc. Lond. A 117, 360 (1978).
[23] A. Linde, Particle Physics and Inflationary Cosmology. Academic Press, San Diego (1990).
[24] U. H. Danielson, Phys. Rev. D 66, 23511 (2002) [hep-th/0203198].
[25] J. Bognar, Indefinite Inner Product Space, Springer-Verlag, (1974).
[26] B. Forghan, M. V. Takook and A. Zarei, Annals of Physics 327, 2388, (2012) [hep-ph/1206.2796].
[27] N. D. Birrel, P. C. W. Davies, Quantum Fields in Curved Space, Cambridge University Press, Cambridge (1982).
[28] M. Abramowitz and I. Stegun, Handbook of Mathematical Functions, with Formulas, Graphs, and Mathematical Tables, Dover (1974).
[29] A. Kempf and J. C. Niemeyer, Phys. Rev. D 64, 103501 (2001) [astro-ph/0103225].
[30] H. Pejhan, M. V. Takook and M. R. Tanhayi, Annals of Physics 341, 195 (2014) [math-ph/1204.6001].